\documentstyle[12pt,epsf,epsfig]{article}
\def\beq{\begin{equation}}   \def\eeq{\end{equation}}
\def\lsim{\mathrel{\rlap{\lower3pt\hbox{\hskip0pt$\sim$}}
    \raise1pt\hbox{$<$}}}         %less than or approx. symbol
\def\gsim{\mathrel{\rlap{\lower4pt\hbox{\hskip1pt$\sim$}}
    \raise1pt\hbox{$>$}}}         %greater than or approx. symbol
\begin{document}
\begin{titlepage}
\def\simlt{\mathrel{\raise.3ex\hbox{$<$\kern-.75em\lower1ex\hbox{$\sim$}}}}
\def\simgt{\mathrel{\raise.3ex\hbox{$>$\kern-.75em\lower1ex\hbox{$\sim$}}}}
\begin{flushright}
TPI--MINN--02/07\\
UMN--TH--2047/02 \\
April 2002
\end{flushright}
\begin{center}
\baselineskip25pt

\vspace{1cm}

{\Large\bf The Forward--Backward Asymmetry of $B\rightarrow (\pi,K)\ell^+\ell^-$:
Supersymmetry at Work}

\vspace{1cm}

{\bf
D. A. Demir, Keith A. Olive,}
\vspace{0.3cm}

Theoretical Physics Institute,
University of Minnesota, Minneapolis, MN 55455
\vspace{0.2cm}

and \\
{\bf M.B. Voloshin}
\vspace{0.3cm}

Theoretical Physics Institute,
University of Minnesota, Minneapolis, MN 55455\\
and\\
Institute of Theoretical and Experimental Physics, Moscow 117218
\end{center}
\vspace{1cm}
\begin{abstract}
We analyze the forward--backward asymmetry of the decays $B\rightarrow (\pi,K)\ell^+\ell^-$
with $\ell=\mu,$ or  $\tau$ in the framework of the constrained  minimal 
supersymmetric standard model. We find that, the asymmetry is enhanced at large
$\tan\beta$ and depends strongly  on the sign of the $\mu$ parameter. For
$\mu>0$, the asymmetry is typically large  and observable whereas for
$\mu<0$, it changes the sign and is suppressed by an order of magnitude. 
Including  cosmological constraints we find that the asymmetry  has a
maximal value of about 30 \%, produced when Higgs- and gauge- induced
flavor violations are of comparable size, at a value of
$\tan\beta \simeq 35$.   The present constraints from 
the $B$--factories are too weak to constrain parameter space, and the regions excluded
by them  are already disfavoured by at least one of $\mbox{BR}(B\rightarrow X_s
\gamma)$, $g-2$, and/or cosmology.  The size of the asymmetry is mainly determined by
the flavor of the final state lepton rather than the flavor of the pseudoscalar.

\end{abstract}

\end{titlepage}

\section{Introduction}
There are sound theoretical and experimental reasons
for studying flavor--changing neutral current 
(FCNC) processes. Such transitions, being forbidden 
at tree level, provide stringent tests of the 
standard model (SM) at the loop level.
Besides, FCNCs form a natural arena for discovering
indirect effects of possible TeV--scale extensions 
of the SM such as supersymmetry. Among all the FCNC phenomena, 
the rare decays of the $B$ mesons are particularly important
as many of the nonperturbative effects are small and under control.
%(See the review \cite{buras}).

In addition to having 
already determined the branching ratio of $B\rightarrow 
X_s \gamma$ \cite{bsgam} and the CP asymmetry of 
$B\rightarrow J/\psi K$ \cite{acp}, experimental activity in $B$ physics, has
begun to probe FCNC phenomena in semileptonic $B$ decays 
\cite{belle,babar,cdf1,cleo1}. Therefore, with increasing data and statistics,
these experiments are expected to give precise measurements on long-- and
short--distance effects in semileptonic decays, $e.g.$  
$B\rightarrow P \ell^+\ell^-$ ($P=K, \pi$) and 
$B\rightarrow V \ell^+\ell^-$ ($V=K^{\ast}, \rho$). 
 The key physical quantities
that can be measured are the branching ratios, 
CP asymmetries and several lepton asymmetries.

In searching for the physics beyond the SM it is often
necessary to deal with quantities that  
differ significantly from their SM counterparts. This is because 
there are large uncertainties coming from the 
hadronic form factors making it hard to disentangle
new physics effects from those of the hadronic dynamics.
For this reason, the pseudoscalar channel $B\rightarrow P \ell^+\ell^-$
provides a unique opportunity, since the forward--backward 
asymmetry $\mbox{A}_{\small \mbox{FB}}$ in this channel 
is extremely small in the SM 
(due to a suppression of order $m_{\ell}m_{b}/M_{W}^{2}$), 
and this remains true in any of its extensions unless a 
scalar--scalar type four--fermion operator such as
\begin{eqnarray}
\label{oper}
\Delta {\cal{A}}&=&\frac{\alpha G_F}{\sqrt{2} \pi}\ V_{tb} V_{tq}^{\ast}\ 
{\cal{C}}(m_b)\ \overline{q_L} b_R \cdot \overline{\ell} \ell  
\end{eqnarray}
provides a significant contribution to the decay amplitude.
Clearly, such operator structures can arise only from the exchange of a scalar 
between the quark and lepton lines with flavor--violating couplings 
to the quarks. For instance, by extending the SM Higgs sector to two SU(2)
doublets, operator structures of the form (\ref{oper}) can be generated 
\cite{notreeFCNC} excluding the possibility of $ad$ $hoc$ tree level FCNCs.
Although the coefficient ${\cal{C}}(m_b)$ in eq.(\ref{oper}) is still proportional
to the lepton mass, it can receive an enhancement when the
ratio of the two Higgs vacuum expectation values, $\tan
\beta$ is large.

Supersymmetry (SUSY), is one of the most favoured extensions of the SM 
which stabilizes the scalar sector against ultraviolet divergences, and
naturally avoids the dangerous tree level FCNC couplings by coupling the
Higgs doublet $H_u$ ($H_d$) to up--type quarks (down--type quarks and 
charged leptons). The soft--breaking of SUSY at the weak scale generates
($i$) a variety of new sources for tree level flavor violation depending
on the structure of the soft terms, and ($ii$) radiatively generates 
various FCNC couplings even if the flavor violation is restricted to
the CKM matrix. The first effect, which cannot be determined theoretically,
is strongly constrained by the FCNC data \cite{masiero}, and therefore,
as a predictive case, it is convenient to restrict all flavor--violating 
transitions to the charged--current interactions where they proceed via 
the known CKM angles. This is indeed the case in various SUSY--breaking 
schemes where hidden sector breaking is transmitted to the observable 
sector via  flavor--blind interactions $e.g.$  gauge--mediated and minimal
gravity--mediated scenarios. This minimal
flavor violation scheme adopted here is well motivated by minimal supergravity
in which all scalars receive a common soft mass, $m_0$, at the
unification scale.

The common origin for scalar masses is one of the parameter restrictions
which define the constrained version of the supersymmetric standard model (CMSSM).
The low energy sparticle spectrum in the CMSSM is specified entirely by 
four parameters, and one sign.  In addition to $m_0$ and $tan \beta$, the
remaining mass parameters are the gaugino masses and supersymmetry breaking
trilinear mass terms.  These too, are assumed to have common values, $m_{1/2}$
and $A_0$, at the  unification scale. In principle, there are two
additional parameters, the Higgs mixing mass, $\mu$, and the
supersymmetry breaking bilinear mass term, $B$, but since it is common to
choose $\tan \beta$ as a free parameter and since we fix the sum of the
squares of the two Higgs vevs with $M_Z$, these two parameters are fixed
by the requirements of low energy electroweak  symmetry breaking. One is
left simply with a sign ambiguity for $\mu$.  Therefore,  the parameters
which define a CMSSM model are $\{ m_{1/2}, m_0, A_0, \tan \beta$ and
$sgn( \mu) \}$.

Although flavor violation is restricted to the CKM matrix, 
radiative effects still generate FCNC transitions among which those
that are enhanced at large values of $\tan\beta$ are particularly important as
the LEP era ended with a clear preference to large values of
$\tan\beta$~\cite{LEPHiggs}.  Indeed, it is known that there are large
$\tan\beta$--enhanced  threshold corrections to the CKM entries
\cite{threshold1} allowing  for Higgs--mediated FCNC transitions
\cite{threshold2}. For instance, the holomorphic mass term for down--type
quarks $H_d Q D^{c} $ acquires a non-holomorphic correction
$H_{u}^{\dagger} Q D^{c}$ where the latter term proportional to
$\tan\beta/16\pi^{2}$, which is not necessarily small at large
$\tan\beta$.

In what follows, we will compute the forward--backward asymmetry of 
$B\rightarrow (\pi, K) \ell^+ \ell^-$ decays in the MSSM.   After
deriving the scalar exchange amplitudes (\ref{oper}), we discuss several
theoretical and experimental issues and then identify the regions of SUSY
parameter space for which the asymmetry is enhanced. We compare our
results with existing experimental and cosmological constraints.

\section{$B\rightarrow (\pi, K) \ell^+ \ell^-$ in Supersymmetry}
In general, the semileptonic decays $B\rightarrow (\pi, K) \ell^+ \ell^-$
proceed via the quark transitions $b\rightarrow (s,d)\ell^+\ell^-$. The 
decay amplitude has the form 
\begin{eqnarray}
\label{amp}
{\cal{A}} &=& \frac{\alpha G_F}{\pi \sqrt{2}}\ V_{tb} V_{tq}^{\star}\ \Bigg[
{\cal{C}}_{7}^{eff}(m_b)\ \overline{q_L} i \sigma_{\mu\nu} k^{\nu} b_R 
\overline{\ell} \gamma^{\mu} \ell + {\cal{C}}_{9}^{eff}(m_b,s)\ \overline{q_L}
\gamma_{\mu} b_{L} \overline{\ell} \gamma^{\mu} \ell\nonumber\\ &+&
{\cal{C}}_{10}(m_b)\ \overline{q_L} \gamma_{\mu} b_{L} \overline{\ell}
\gamma^{\mu} \gamma_{5} \ell   + {\cal{C}}(m_b)\ \overline{q_L} b_{R}
\overline{\ell} \ell  +\widehat{{\cal{C}}}(m_b)\ \overline{q_L} b_{R}
\overline{\ell} \gamma_5 \ell \Bigg]
\end{eqnarray}
where $k_{\nu}=- (2 m_b/q^2) q_{\nu}$ with $q^2\equiv s M_B^2$ being the 
dilepton invariant mass. The Wilson coefficients, ${\cal{C}}_{7},
{\cal{C}}_{9}$, and ${\cal{C}}_{10}$  have been computed to leading
order in \cite{borzumati}. Higher order
${\cal{O}}(\alpha_s)$ corrections, which are available for small $s$ in
the SM \cite{asatryan}, will not be considered. The coefficients
${\cal{C}}$ and $\widehat{{\cal{C}}}$ will be discussed below.

The kinematical range for the normalized dilepton invariant mass in terms of
the lepton and pseudo scalar masses is 
$4 m_{\ell}^{2}/M_{B}^{2} \leq s \leq (1-M_P/M_{B})^{2}$ which
includes the vector charmonium resonances $J/\psi, \psi^{\prime}, \psi^{\prime\prime},\cdots$
whose effects are included in the ${\cal{C}}_{9}^{eff}(m_b,s)$. Moreover, 
the four--fermion operators for the light quarks develop nonvanishing
matrix elements, and these are also included in ${\cal{C}}_{9}^{eff}(m_b,s)$.
At higher orders in $\alpha_s$, these effects contribute to ${\cal{C}}_{7}^{eff}(m_b)$ 
as well \cite{asatryan}.

The electromagnetic dipole coefficient ${\cal{C}}_{7}^{eff}(m_b)$ is contributed
by graphs with the W boson, charged Higgs, and chargino penguins. The chargino
contribution  increases linearly with $\tan\beta$ at  leading order
\cite{borzumati}, and the inclusion of SUSY threshold corrections strengthens
this dependence \cite{bsgam2}. This coefficient is directly constrained by the
$B\rightarrow X_s \gamma$  decay rate, and the experimental bounds can be
satisfied with a relatively  light charged Higgs at very large values of
$\tan\beta$. On the other hand, the coefficient of the vector--vector
operator
${\cal{C}}_{9}^{eff}(m_b,s)$ is generated by box diagrams, and carries a
long--distance piece  coming from the matrix elements of the light quark
operators as well as the intermediate charmonium states \cite{kruger}. Finally,
the  coefficient of the vector--pseudovector operator ${\cal{C}}_{10}(m_b)$
is generated by box graphs and is scale independent. Both 
coefficients ${\cal{C}}_{9}^{eff}(m_b,s)$ and ${\cal{C}}_{10}(m_b)$
are less sensitive to $\tan\beta$ than is ${\cal{C}}_{7}^{eff}(m_b)$.

Within the SM, these coefficients typically have the values ${\cal{C}}_{7}^{eff}(m_b)
\approx -0.3$, ${\cal{C}}_{9}^{eff}(m_b,s)\approx 4.4$ (excluding its 
long--distance part), and ${\cal{C}}_{10}(m_b)\approx -4.7$ \cite{kruger}
which, however, are allowed to vary considerably within the existing bounds 
\cite{biz}. The inclusion of SUSY contributions, for instance, implies large 
variations in ${\cal{C}}_{7}^{eff}(m_b)$ (even changing its sign),  
and typically a $\sim 10\%$ variation in ${\cal{C}}_{9}^{eff}(m_b,s)$ 
and ${\cal{C}}_{10}(m_b)$ \cite{susyeff}.

The scalar--scalar operators in the decay amplitude are generically 
induced by the exchange of the Higgs scalars and suffer invariably 
from the $m_{\ell} m_{b}/M_{W}^{2}$ suppression. Therefore, these
operators are completely negligible in the SM. However, in the
MSSM, this suppression is overcome by large $\tan\beta$
effects where the charged Higgs--top diagram is proportional to
$\tan^{2}\beta$,  and the chargino--stop diagram is $\sim {\cal{O}}(\tan^{3}\beta)$. In more
explicit terms, $\widehat{{\cal{C}}}(m_b)=-{\cal{C}}_{10}(m_b)$, and 
\begin{eqnarray}
\label{cmb}
{\cal{C}}(m_b)&=&\frac{ 2 m_b m_{\ell} G_F}{\sqrt{2}}\ \frac{1}{4 \pi
\alpha}\ 
\frac{1}{(1+\epsilon_{g} \tan\beta)(1+(\epsilon_g+h_t^{2} \epsilon_{h})\tan\beta)}\ 
\nonumber\\&\times& \left[ \tan^{2}\beta f(x_{tH}) - \epsilon_{\mu} \tan^{3}\beta  x_{tA} 
\frac{M_{\chi^{\pm}} A_t} {M_{\tilde{t}_1}^{2}-M_{\chi^{\pm}}^{2}}\ 
\left(f(x_{\chi^{\pm} \tilde{t}_2})-f(x_{\tilde{t}_1
\tilde{t}_2})\right)\right]
\end{eqnarray}
where $\epsilon_{\mu}$ is the sign of the $\mu$ parameter, $\chi^{\pm}$ is
the  lighter chargino, $x_{ij}=m_{i}^{2}/m_{j}^{2}$, $f(x)=x \log{x}
/(1-x)$, and the parameters $\epsilon_{g}$ and $\epsilon_{h}$, which are
typically ${\cal{O}}(10^{-2})$, are defined in \cite{threshold2,bsgam2}.
Finally, $h_t$ is the top quark Yukawa coupling, $M_{\tilde{t}_1}$ is the
light stop mass, and $A_t$ is the low energy value of the SUSY breaking
top-Yukawa trilinear mass term obtained from $A_0$ by the running of the
RGEs.  Clearly, the charged Higgs contribution,  which is the dominant one
in two--doublet models
\cite{notreeFCNC}, is subleading compared to the chargino contribution.
The sign of
${\cal{C}}(m_b)$ depends explicitly on
$\epsilon_\mu$. Therefore, the  forward--backward asymmetry in
$B\rightarrow (\pi,K)\ell^+\ell^-$ decays depends strongly on the sign of
the $\mu$ parameter.

{}From the experimental point of view, it is useful to analyze the 
normalized forward--backward asymmetry defined as 
\begin{eqnarray}
\mbox{A}_{\mbox{\small FB}}(P\ell^+\ell^-)=\frac{\int_{-1}^{0} dz \frac{d^{2} \Gamma}{dz ds}
-\int_{0}^{1} dz \frac{d^{2} \Gamma}{dz ds}}{\int_{-1}^{0} dz \frac{d^{2} \Gamma}{dz ds}
+\int_{0}^{1} dz \frac{d^{2} \Gamma}{dz ds}}
\end{eqnarray}
where $z=\cos \theta$, $\theta$ being the angle between the momenta 
of $P$ and $\ell^{+}$. A direct calculation gives the explicit expression
\begin{eqnarray}
\label{afb}
\mbox{A}_{\mbox{\small FB}}(P\ell^+\ell^-)=-\frac{\lambda^{1/2}(s) v(s) t_{\ell} \mbox{Re}[A_{79}(s)] 
A(s)}{\Sigma(s)}
\end{eqnarray}
where 
\begin{eqnarray}
\Sigma(s)&=&\lambda(s) (1- v(s)^2/3) \left[|A_{79}(s)|^{2}+A_{10}(s)^{2}\right] +
t_{\ell} t_{P} A_{10}(s)^{2}\nonumber\\ &+& s
t_{\ell}\left[(B_{10}(s)-A(s))^{2}+v(s)^{2}  A(s)^{2}\right]\nonumber\\&+&
2 t_{\ell} (1-t_{P}/4-s) [B_{10}(s)-A(s)] A_{10}(s)\;.
\end{eqnarray}
Here $t_{\ell}= 4 m_{\ell}^{2}/M_{B}^{2}$, $t_{P}=4 M_{P}^{2}/M_{B}^{2}$, 
$\lambda(s)=(1-s-t_P/4)^{2}-s t_{P}$, $v(s) = (1-t_\ell/s)^{1/2}$ and 
\begin{eqnarray} 
A_{79}(s)&=&{\cal{C}}_{9}^{eff}(m_b,s)f_{+}(s)-{\cal{C}}_{7}^{eff}(m_b)f_{7}(s)\,,\nonumber\\
A_{10}(s)&=&{\cal{C}}_{10}(m_b) f_{+}(s)\;\:\:,\:\:\: B_{10}(s)={\cal{C}}_{10}(m_b) 
(f_{+}(s) + f_{-}(s))\nonumber\\
A(s)&=& {\cal{C}}(m_b)\ \frac{M_B^2}{2 m_b m_{\ell}}\ [(1-M_{P}^2/M_{B}^2)f_{+}(s)+s f_{-}(s)]
\end{eqnarray}
where $f_{7}(s)=(2 m_b)/(M_{B}+M_{P}) f_{T}(s)$.  
The form factors  $f_{+}$, $f_{-}$ and $f_{T}$ are not 
measured at present and one has to rely on theoretical predictions. 
In what follows we use the results of the calculation \cite{ball} of 
these form factors from QCD sum rules for both $B\rightarrow K$ and 
$B\rightarrow \pi$ transitions.

In general, the hadronic form factors are uncertain by $\sim 15\%$,
and this translates into an uncertainty of approximately $35\%$ in
the branching ratio. Especially for low $s$, below the charmonium resonances, 
the theoretical prediction for the branching ratio contains
large uncertainties \cite{ali}. Therefore, theoretically the 
large dilepton mass region is more tractable. On top of the
form factor uncertainties, there are further problems in treating 
the contributions of the charmonium resonances (embedded 
in the Wilson coefficient ${\cal{C}}_{9}^{eff}(m_b,s)$). For instance,
the recent BELLE experiments \cite{belle}, subtract such resonance contributions
by vetoing the range $0.322\lsim s \lsim 0.362$. Then the 
experimental bound on the branching ratio turns out to be 
\begin{eqnarray}
\label{belle}
0.0328\leq 10^{6}\times \mbox{BR}(B\rightarrow K \mu^+\mu^-)\leq 2.395\:\:\: \mbox{at}\:\: 90\% \: \mbox{C.L.}
\end{eqnarray}
which we will take into account in making the numerical estimates
below. It can be also noted that the same decay mode has not been observed
by BABAR: $\mbox{BR}(B\rightarrow K \mu^+\mu^-)< 4.5 \times 10^{-6}\:\: \mbox{at}\:\: 
90\% \: \mbox{C.L.}$ \cite{babar}. In addition,  the vector kaon final states have not
been observed yet: $\mbox{BR}(B\rightarrow K^{\star} 
\mu^+\mu^-)< 3.6 \times 10^{-6}\:\: \mbox{at}\:\: 
90\% \: \mbox{C.L.} $ \cite{belle,babar}. One notes that,
the asymmetry is large in regions of the parameter space where 
the branching ratio is depleted, and therefore, the BELLE
lower bound on $B\rightarrow K \mu^+\mu^-$ is an important constraint 
which can prohibit the asymmetry taking large marginal values. 
Clearly, in the presence of ${\cal{C}}(m_b)$, which can take 
large values in SUSY, the would--be experimental constraints on the 
$C_{10}(m_b)$--${\cal{C}}_{9}^{eff}(m_b,s)$ plane are  lifted.

Furthermore, the pure leptonic decay modes, $B_{s,d}\rightarrow \ell^+\ell^-$,
depend directly on the Wilson coefficients $C_{10}(m_b)$, ${\cal{C}}(m_b)$ and
$\widehat{{\cal{C}}}(m_b)$. In the SM, $\mbox{BR}(B_{s}\rightarrow
\mu^+\mu^-)\approx 10^{-9}$ which is approximately three orders of magnitude
below the present bounds $\mbox{BR}(B_{s}\rightarrow \mu^+\mu^-)<  2.6\times
10^{-6}$ \cite{cdf}. The SUSY contributions, especially at large $\tan\beta$,
can  enhance the SM prediction typically by an order of magnitude, and the
bounds can even be violated in certain corners of the parameter space
\cite{bsll}. In what follows the constraints from $B_{s}\rightarrow \mu^+\mu^-$
as well as the muon $g-2$ (as they are directly correlated \cite{nierste}) will
be taken into account. We will refer to the constraints from $B\rightarrow
(K,K^{\star})\mu^+\mu^-$ and 
$B_{s,d}\rightarrow \ell^+\ell^-$ collectively as $B$--factory
constraints.

\begin{figure}
%\vspace*{-0.75in}
\hspace*{-0.7in}
\begin{minipage}{7.5in}
\begin{center}
\epsfig{file=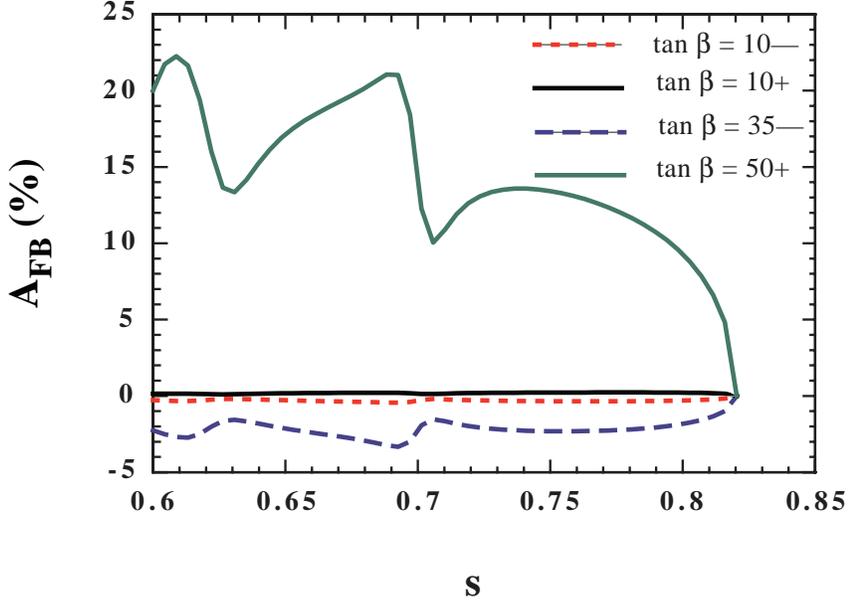,height=3.25in}
\end{center}
\end{minipage}
\vskip .2in
\caption{\label{fig6}
{\it The dependence of $\mbox{A}_{\mbox{\small FB}}(K\tau^+\tau^-)$ 
on the normalized dilepton invariant mass, $s=q^2/M_B^2$, for various (allowed)
points  in the SUSY parameter space. Each curve is labeled as $\tan\beta\
\epsilon_{\mu}$ where $(m_{1/2},m_0)=(400,100),(400,100),(900,700),(1200,600)\
{\rm GeV}$ from top to bottom. One notices that the
asymmetry is typically small for $\mu<0$ as was noted at various instances
before.}}
\end{figure}

For the constraints on the SUSY parameter space to make sense it is 
necessary to be far from the regions of large hadronic uncertainties, and
thus,  below we will restrict the range of $s$ to lie well above the
charmonium  resonances and well below the kinematical end point. 
In Fig. \ref{fig6}, we show the variation of the asymmetry with the
normalized dilepton invariant mass, $s=q^2/M_B^2$ for various 
values of the SUSY parameters (see below for further discussion of these
choices).  The irregularities in the $s$--dependence of the
asymmetry  are similar to those in the $B\rightarrow K^{\star}
\ell^+\ell^-$ decay. The various bumps and valleys come from
the relative sizes of individual terms contributing to  
$\mbox{A}_{\mbox{\small FB}}$.  It should be noted that in the region
around the value $s=0.75$,  the $s$
dependence of the asymmetry is rather smooth.
Therefore,
in forming the constant asymmetry contours in the space of SUSY parameters
we will take $s=0.75$ (corresponding to $q^{2}=20.1\ {\rm GeV}^{2}$).

At low values of the asymmetry one should also take into account
the final state electromagnetic interactions. Indeed, photon exchange
between the lepton and $P=K,\pi$ lines is expected to induce an
asymmetry ${\cal{O}}(\alpha/\pi)$ implying that only asymmetries
$\mbox{A}_{\mbox{\small FB}}$ larger than $\sim 1\%$ can be trusted to 
follow from SUSY effects, unless the interplay with the electromagnetic 
corrections is explicitly taken into account. For large values of
asymmetry, when the observation of the effect becomes feasible, higher
order QCD effects (not yet calculated) can in principle modify our
results somewhat. However, it is highly unlikely that these corrections
will dramatically reduce the asymmetry discussed here.

One should also note that in the limit of exact $SU(3)$ flavor symmetry,
the asymmetries in $B\rightarrow K \ell^+\ell^-$ and $B\rightarrow \pi \ell^+\ell^-$
decays must be the same. Due to $SU(3)$ breaking effects, which
show up in different parameterizations of the form
factors $f_{+}$, $f_{-}$ and $f_7$ for $B\rightarrow \pi$ and
$B\rightarrow K$ transitions \cite{ball}, their asymmetries are expected
to differ slightly. Clearly it is the lepton flavor that largely
determines the size of the asymmetry, rather than the flavor of the final
state pseudoscalar.

Finally, before starting to scan the SUSY parameter space, it is worth while
discussing the sensitivity of $\mbox{A}_{\mbox{\small FB}}$ to some of
the parameters. First of all, as $\tan\beta$ grows, two of the Wilson
coefficients, ${\cal{C}}_{7}^{eff}(m_b)$ and ${\cal{C}}(m_b)$, grow
rapidly up to the bounds obtained from rates of the decays 
$(B\rightarrow X_s \gamma)$ and $(B\rightarrow 
K \mu^+\mu^-)$. Since $\mbox{Re}[{\cal{C}}_{9}^{eff}(m_b,s)] >0$ and
${\cal{C}}_{7}^{eff}(m_b)>0$,
$\mbox{Re}[A_{79}(s)]$ increases with $\tan\beta$. However, this
increase is much milder than the $\tan^{3}\beta$ dependence of ${\cal{C}}(m_b)$
causing $A(s)$ to take large negative (positive) values for $\mu>0$ ($\mu<0$).
Therefore, large $\tan\beta$ effects influence not only the 
numerator of Eq. (\ref{afb}) but also the denominator $\Sigma(s)$ (proportional
to the differential branching fraction) via the destructive (constructive)
interference with $B_{10}(s)$ ($C_{10}(m_b)$ remains negative in
SUSY) for $\mu>0$ ($\mu<0$). However, as $\tan\beta$ keeps growing,
depending on the rest of the SUSY parameters, the effect of ${\cal{C}}(m_b)$
eventually becomes more important, and the asymmetry falls 
rapidly due to the enhanced branching ratio. In this sense, the regions of
enhanced  asymmetry depend crucially on the sign of the $\mu$ parameter 
and the specific value of $\tan\beta$. Moreover, as the expression of
${\cal{C}}(m_b)$ makes clear, there can be sign changes 
in the asymmetry in certain regions of the parameter 
space due to the relative sizes of the masses
of the lighter chargino and stops. Such effects will also
give small asymmetries just like the $\mu<0$ case.

Our work extends a
previous analysis of this asymmetry \cite{kruger2} by
including
the large gluino exchange effects (contained in the 
quantity $\epsilon_g$ in  Eq. (\ref{cmb})) and the explicit 
dependence on the sign of the $\mu$ parameter. In addition,
we go beyond the work 
in \cite{kruger2} as well as in the preceding work 
\cite{chinese} by resumming the higher order $\tan\beta$ terms 
which increases the validity of 
the analysis at large values of $\tan\beta$ \cite{bsgam2}.
We note that a computation of the large $\tan\beta$
effects can be carried out in the gaugeless limit \cite{threshold2}
which eliminates some of the diagrams considered in \cite{kruger2}.

In the numerical analysis below, we will analyze the forward--backward 
asymmetry of the decays $B\rightarrow P \ell^+\ell^-$ by taking into 
account the above--mentioned constraints from $B$ factories as well as
other collider and cosmological constraints. We will be searching for those
regions of the SUSY parameter space in which the asymmetry is enhanced. In
particular, we will be particularly  interested in the sensitivity of the
asymmetry to $\tan\beta$, the sign of  the $\mu$ parameter, as well as the
common scalar mass $m_0$ and the gaugino mass $m_{1/2}$.

\section{Results}

In our analysis, we include several accelerator as well as cosmological
constraints.  From the chargino searches at 
LEP \cite{LEPsusy}, we
apply the kinematical 
limit $m_{\chi^\pm} \gsim$ 104 GeV.  A more careful consideration of the
constraint would lead to an unobservable difference in the figures shown below. 
This constraint can be translated into a lower bound on the gaugino mass
parameter, $m_{1/2}$ and is nearly independent of other SUSY parameters.  The LEP
chargino limit is generally overshadowed (in the CMSSM) by the important
constraint provided by the LEP lower limit on the Higgs mass:
$m_h > $ 114.1 GeV \cite{LEPHiggs}. This holds in the Standard Model, for
the lightest Higgs boson $h$ in the general MSSM for $\tan\beta
\lsim 8$, and almost always in the CMSSM for all $\tan\beta$. 
The Higgs limit also imposes important constraints on the CMSSM parameters,
principally $m_{1/2}$,  though in this case, there is a strong
dependence on $\tan \beta$. The Higgs masses 
are  calculated here using {\tt FeynHiggs}~\cite{FeynHiggs}, which is estimated to 
have a residual uncertainty of a couple of GeV in $m_h$.

We also include  the constraint imposed by
measurements of $b\rightarrow s\gamma$~\cite{bsgam,bsgam2}. These agree
with the Standard Model, and therefore provide bounds on MSSM particles, 
such as the chargino and charged Higgs
masses, in particular. Typically, the $b\rightarrow s\gamma$
constraint is more important for $\mu < 0$, but it is also relevant for $\mu >
0$,  particularly when $\tan\beta$ is large.

The final experimental constraint we consider is that due to the
measurement of the anomalous magnetic moment of the muon.  The BNL E821~\cite{BNL}
experiment reported a new measurement of $a_\mu\equiv {1\over 2} (g_\mu -2)$
which deviates by 1.6 standard deviations from the best Standard Model prediction
(once the pseudoscalar-meson pole part of the light-by-light 
scattering contribution \cite{lightbylight} is corrected).  Although negative
values of $\mu$ are no longer entirely excluded~\cite{susygmu}, the 2-$\sigma$
limit still excludes much of the $\mu < 0$ parameter space \cite{EOS2}. 
$\mu < 0 $ is allowed so long as either (or both) $m_{1/2}$
and $m_0$ are large.

We also apply the cosmological limit on the
relic density of the lightest supersymmetric particle (LSP), $\rho_\chi =
\Omega_\chi
\rho_{critical}$, and require that
\beq
0.1 < \Omega_\chi h^2 < 0.3
\label{ten}
\eeq
The upper limit is rigorous, and assumes only that the age of the Universe
exceeds 12 Gyr.  It is also consistent with the total matter density $\Omega_m
\lsim 0.4$, and the Hubble expansion rate $h \sim 0.7$ to within about 10
\% (in units of 100 km/s/Mpc). On the other hand, the lower limit in (\ref{ten})
is optional, since there could be other important contributions to the overall
matter density.

The cosmologically allowed regions in the CMSSM have been well studied
\cite{EFGOSi,other}.  There are generally large, `bulk' regions of
parameter space at low to moderate values of $m_{1/2}$ and $m_0$ at all
values of $\tan \beta$. There are additional regions which span out to
large values of $m_{1/2}$ due to co-annihilations with light sleptons,
particularly the lighter $\tilde \tau$
\cite{ourcoann}.  At large $\tan \beta$, there are also regions in which
the lightest neutralino sits on the s-channel pole of the pseudo-scalar and heavy
scalar Higgs producing `funnel'-like regions \cite{funnel,EFGOSi}. Finally, 
there are the so-called `focus-point' regions \cite{focus} which are present at
very large values of $m_0$.  Generally, these regions have a lower
asymmetry (because of the large value of $m_0$), however, at values of
$\tan \beta \sim 50$, asymmetries as large as 10\% are possible.

\begin{figure}
\vspace*{-0.25in}
%\hspace*{-0.7in}
\begin{minipage}{8in}
\epsfig{file=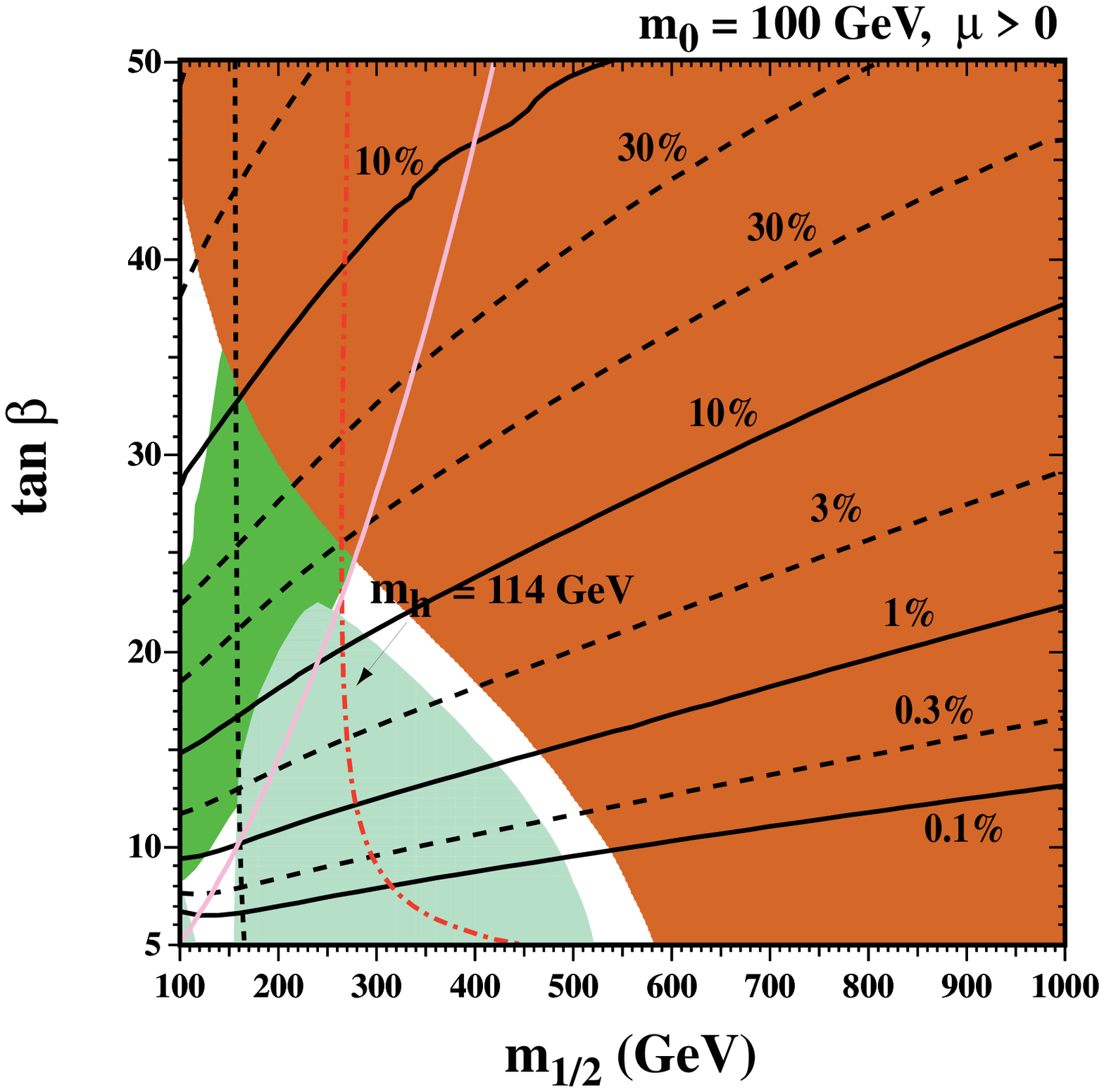,height=3.25in}
\epsfig{file=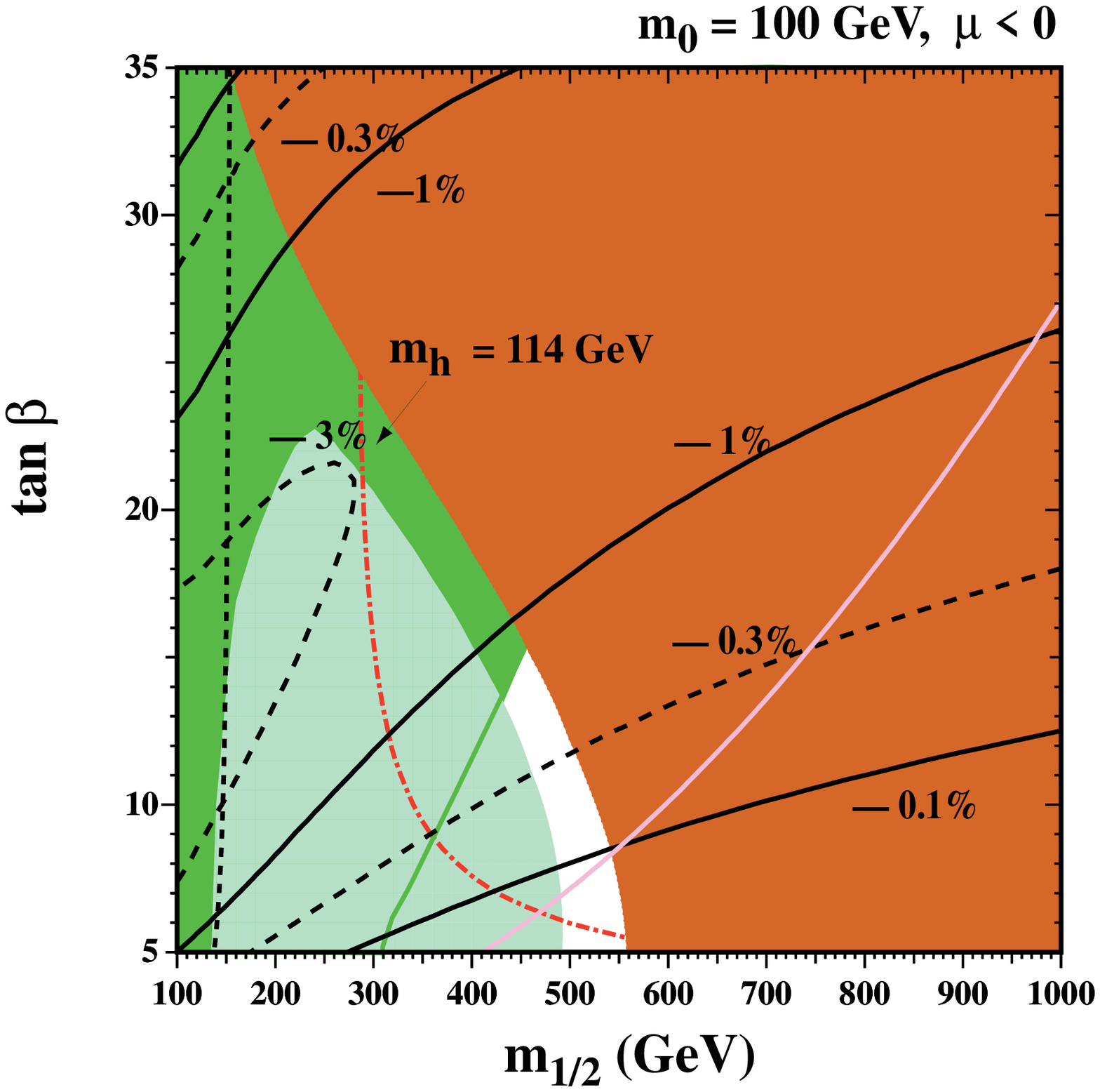,height=3.25in}
\end{minipage}
\begin{minipage}{7.5in}
\epsfig{file=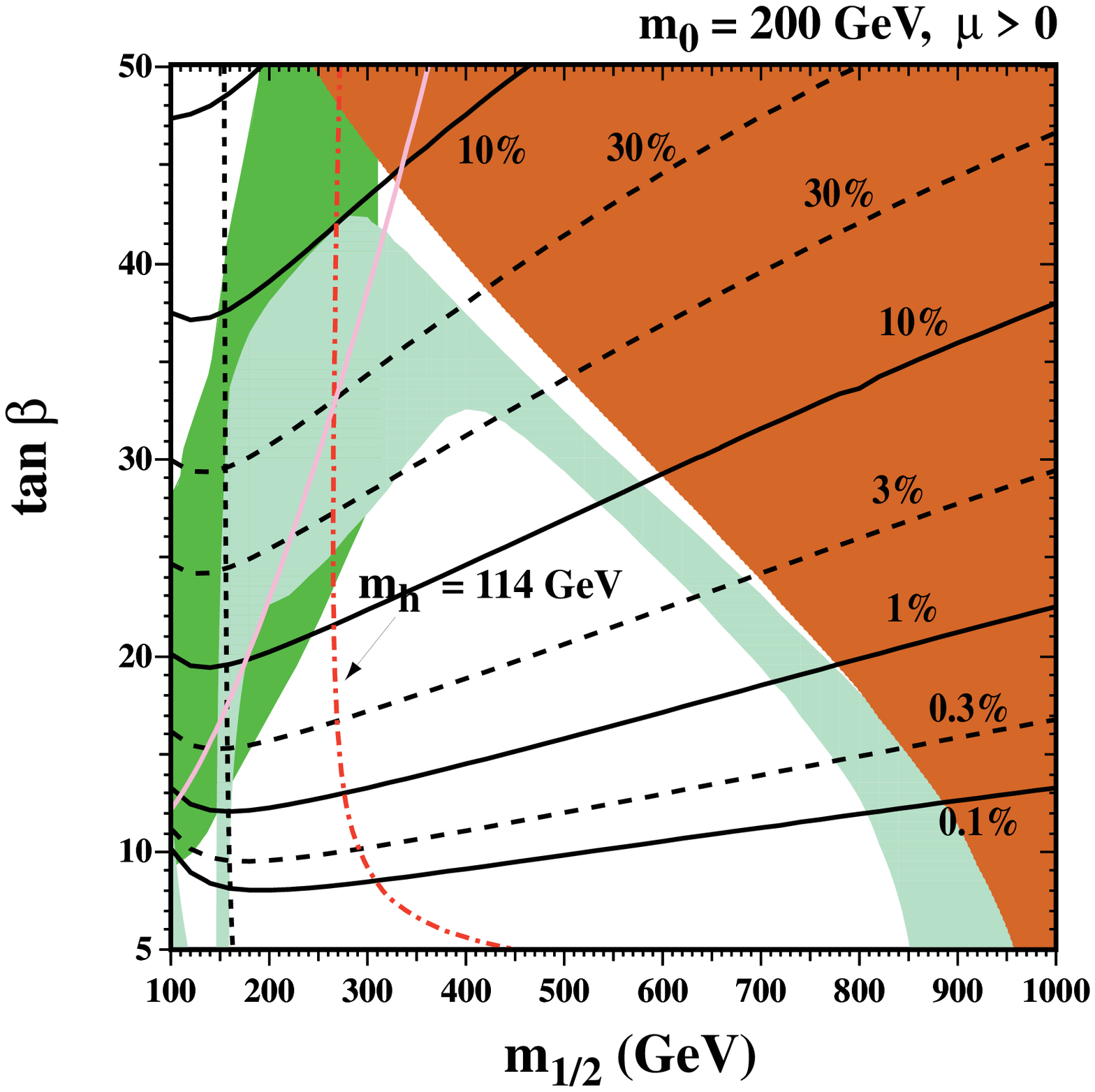,height=3.25in}
\epsfig{file=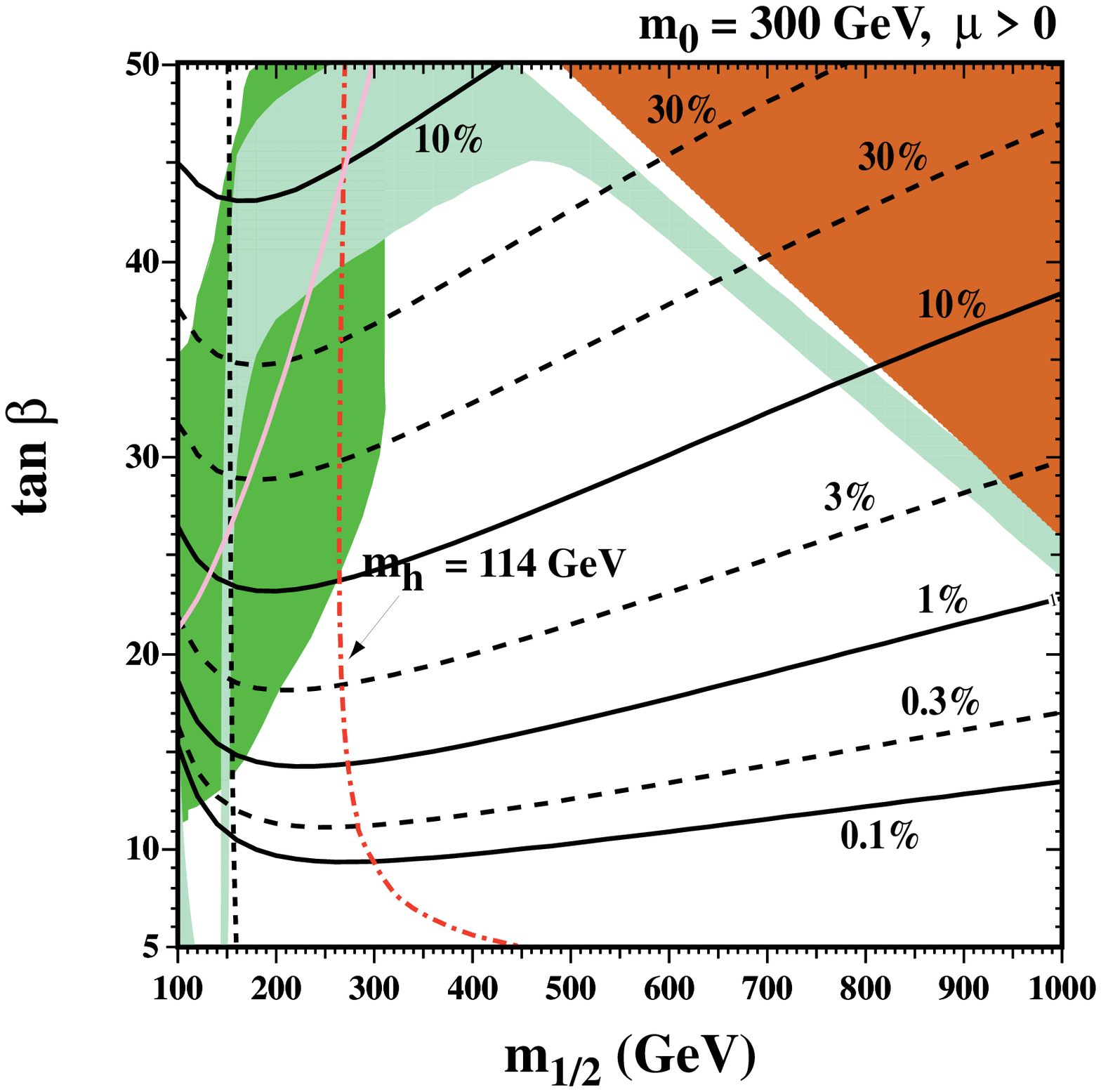,height=3.25in}
\end{minipage}
\vskip .2in
\caption{\label{fig1}
{\it The constant $\mbox{A}_{\mbox{\small FB}}(K\tau^+\tau^-)$
contours in $m_{1/2}$--$\tan\beta$ plane for $A_0=0$, with  $m_{0}=100$ GeV, 
$\mu>0$ (a);  $m_{0}=100$ GeV, $\mu<0$ (b); $m_{0}=200$ GeV, 
$\mu>0$ (c); and  $m_{0}=300$ GeV, $\mu>0$ (d). We take
$m_t=175\ {\rm GeV}$ and $m_b(m_b)_{SM}^{\overline{MS}}=4.25\ {\rm GeV}$. 
In all of the panels the black dashed line shows the 104 GeV chargino mass
contour,  the dot--dashed (red) curve stands for $m_h=114\ {\rm GeV}$
(evaluated using the FeynHiggs code \protect\cite{FeynHiggs}). The light
(turquoise) areas are the cosmologically preferred regions with $0.1\leq
\Omega_{\chi} h^2 \leq 0.3$. In the dark (brick--red) shaded areas the LSP is
$\widetilde{\tau}_1$ and thus excluded.
The medium (green) shaded regions are excluded by $\mbox{BR}(B\rightarrow
X_s
\gamma)$. The  light (violet) curve shows the $g_{\mu}-2$ (the area to the
right of which is allowed).
}}
\end{figure}

In Fig. \ref{fig1}a, we show the contours of constant $\mbox{A}_{\mbox{\small
FB}}(K\tau^+\tau^-)$ in $m_{1/2}$--$\tan\beta$ plane for $A_0=0$, 
$m_{0}=100\ {\rm GeV}$ and $\mu>0$. The constraints discussed above are shown by
various curves and shaded regions. The nearly vertical dashed (black)
line at the left of the figure shows the chargino mass constraint. Allowed
regions are to the right of this line.  The dot-dashed Higgs mass contour
(red) labeled 114 GeV, always provides a stronger constraint. 
Allowed regions are again to the right of this curve.  However, one
should be aware that there is a theoretical uncertainty in the Higgs mass
calculation, making this limit somewhat fuzzy. The light (violet) solid
curve shows the position of the 2-$\sigma$ $g-2$ constraint, which again
excludes small values of $m_{1/2}$.  In the dark (red) shaded region
covering much of the upper left half of the plane, the lighter $\tilde
\tau$ is either the LSP or is tachyonic. Since there are very strong
constraints forbidding charged dark matter, this region is excluded. The
medium shaded (green) region shows the exclusion area provided by the $b
\to s \gamma$ measurements. Finally the light (turquoise) shaded region
shows the area {\it preferred} by cosmology. Outside this shaded region,
the relic density is too small and is technically not excluded.

Putting all of the constraints together, we find that for this value of
$m_0 = 100$~ GeV and $\mu>0$, the allowed region is bounded by 
$300\ {\rm GeV} \lsim m_{1/2}\lsim 500\ {\rm GeV}$ and $5\lsim \tan\beta \lsim
20$. In the allowed region, the forward-backward asymmetry varies rapidly from
very small (unobservable) values  up to $10\%$. There is a wide region
with an observable $1$--$10\%$ asymmetry though the $10\%$ region is quite
narrow (restricted to $\tan \beta \sim 20$).

In Fig. \ref{fig1}b, we show the corresponding result for the opposite sign of
$\mu$. While the cosmologically allowed region is qualitatively similar to the
$\mu > 0$ case and the Higgs limit is slightly stronger, we see that the
$b \to s \gamma$ constraint is significantly stronger. Indeed, the
combined constraints from $b \to s \gamma$ and a $\tilde \tau$ LSP
exclude $\tan \beta \gsim 15$ for this value of $m_0$ and $\mu < 0$. The
2-$\sigma$ constraint from $g-2$ is also significantly stronger and when
combined with the $\tilde \tau$ LSP constraint now exclude values of
$\tan \beta \gsim 8$.  
 
  As mentioned earlier, for $\mu < 0$ both the sign and size of the asymmetry have
changed. In general, the size of the asymmetry is suppressed by an order of
magnitude. Clearly, the $b \to s \gamma$ constraint now allows only a small region
with a $-0.3$ -- $-1\%$ asymmetry. However, when all constraints are
combined they  exclude almost completely the otherwise allowed regions.
At higher values of $m_0$, slightly larger asymmetries are possible.
At $m_0 = 200$ GeV (with $\mu < 0$), $b \to s \gamma$ allows asymmetries
as large as $-2 $\%, however, the $g-2$ data still restricts the asymmetry
to values below about $-0.4$ \%. Even at large $\tan \beta$ and very large
$m_0$, we will see below that for $\mu < 0$, asymmetries never excced
$\sim -1$
\%. We note that independent of the sign of
$\mu$, the asymmetry is maximized for intermediate values of
$\tan\beta$, $i.e.$ it does not monotonically increase with increasing
$\tan\beta$ as was already argued earlier.  The main conclusion from this figure
is that the sign of the $\mu$ parameter  must be positive in order to have large
observable $\mbox{A}_{\mbox{\small FB}}(K\tau^+\tau^-)$.

 We note that there are already
$B$--factory constraints due to recent BELLE and BABAR experiments
\cite{acp,belle,babar}. For
$\mu>0$, they exclude a small region (not plotted) with $m_{1/2}\lsim 120\
{\rm GeV}$ and
$\tan\beta\gsim 43.5$ (lying in the region with a charged LSP), whereas for
$\mu<0$ the excluded region is shifted to $m_{1/2}\lsim 260\ {\rm GeV}$ and
$\tan\beta\gsim 30.5$ (now lying in the region also excluded by
$\mbox{BR}(B\rightarrow X_s \gamma)$).

%\begin{figure}
%\vspace*{-0.75in}
%\hspace*{-0.7in}
%\begin{minipage}{7.5in}
%\epsfig{file=asym1c.eps,height=3.25in}
%\epsfig{file=asym1d.eps,height=3.25in}
%\end{minipage}
%\vskip .2in
%\caption{\label{fig2}
%{\it The constant $\mbox{A}_{\mbox{\small FB}}(K\tau^+\tau^-)$ contours
%in $m_{1/2}$--$\tan\beta$ plane for $A_0=0$, $m_0=200\ {\rm GeV}$
%and $\mu>0$.}}
%\end{figure}  

The behaviour observed in Fig. \ref{fig1}a) and b) is subject to large variations
once the GUT--scale input parameters are varied.
%the sparticle contributions start decoupling and thus
%$\mbox{BR}(B\rightarrow X_s \gamma)$ and other measured quantities
%are expected allow for larger values of $\tan\beta$.
This is seen in Fig. \ref{fig1}c) and d) where the constant asymmetry curves
are plotted in $m_{1/2}$--$\tan\beta$ for $\mu>0$, $A_0=0$, and $m_0=200\
{\rm GeV}$ and $m_0=300\ {\rm GeV}$ respectively.
It is clear that with increasing $m_0$ the cosmologically preferred
`bulk' region is shifted towards larger $\tan\beta$ values making it
possible to get larger asymmetries. In addition, we see very clearly the 
effect of $\chi - {\tilde \tau}$ coannihilations \cite{ourcoann} which
extend the cosmological region to high values of $m_{1/2}$. The region
below the `bulk' and coannihilation region is excluded as it corresponds
to an area with 
$\Omega h^2 > 0.3$. While the chargino, Higgs, and $g-2$ constraints are 
only slightly altered at the higher value of $m_0$, we see that the
charged LSP constraint is relaxed in Fig. \ref{fig1}c) and greatly relaxed in
panel d).

For the higher values of $m_0$ we see from Fig. \ref{fig1}c)
 that for $300\ {\rm GeV}\lsim m_{1/2}\lsim
800\ {\rm GeV}$ and $25\lsim \tan\beta\lsim 40$ the forward-backward asymmetry   
ranges from $1\%$ to $30\%$ in regions which are not excluded by any
experimental or cosmological constraints. In particular, when $350\ {\rm
GeV}\lsim m_{1/2}\lsim 400\ {\rm GeV}$ and $30\lsim \tan\beta\lsim 35$ the
asymmetry is well observable with a typical $\sim 30\%$ peak value. Here the
$B$--factory constraints are effective for $m_{1/2}\sim 100\ {\rm GeV}$
and $\tan\beta \gsim 47.5$, i.e. only a small region in the upper left corner.

For the $m_0=300\ {\rm GeV}$ case shown in Fig. \ref{fig1}d), the cosmologically
allowed region is now shifted up past the maximum of 
$\mbox{A}_{\mbox{\small FB}}(K\tau\tau)$, and is now typically $10\%$.
Overall the forward-backward asymmetry is larger than 1\% in the cosmologically
allowed region  and extends over the range
$300\ {\rm GeV} \lsim m_{1/2} \lsim 1000\ {\rm GeV}$
and $25\lsim \tan\beta\lsim 50$.

 We have also checked  some cases with nonzero values of 
$A_0$, assuming it to be either constant ($e.g.$ set to $2\ {\rm TeV}$), or
varying in proportion with $m_{1/2}$ ($e.g.$, $A_0=2 m_{1/2}$). For a
variable $A_0$, results were not qualitatively different from the results
shown here. For a large and fixed value of $A_0$, the comsological
regions of interest could be very different \cite{EOS1}, however, the
asymmetry was found to be quantitatively similar as the results shown
here. However, we can not claim to have made a systematic examination of
the $A_0 \ne 0 $ parameter space.

\begin{figure}
%\vspace*{-0.75in}
%\hspace*{-0.7in}
\begin{minipage}{7.5in}
\epsfig{file=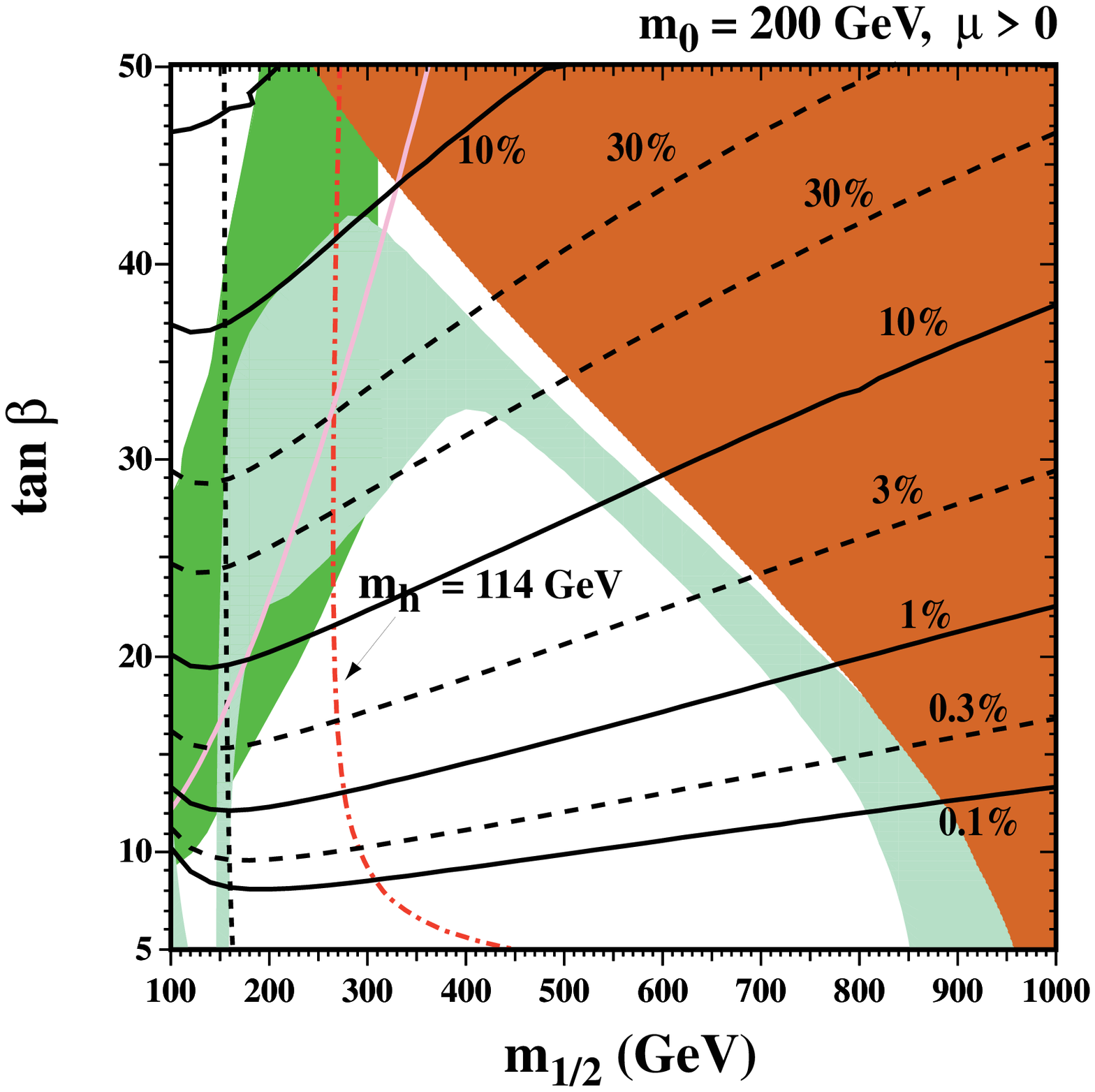 ,height=3.25in}
\epsfig{file=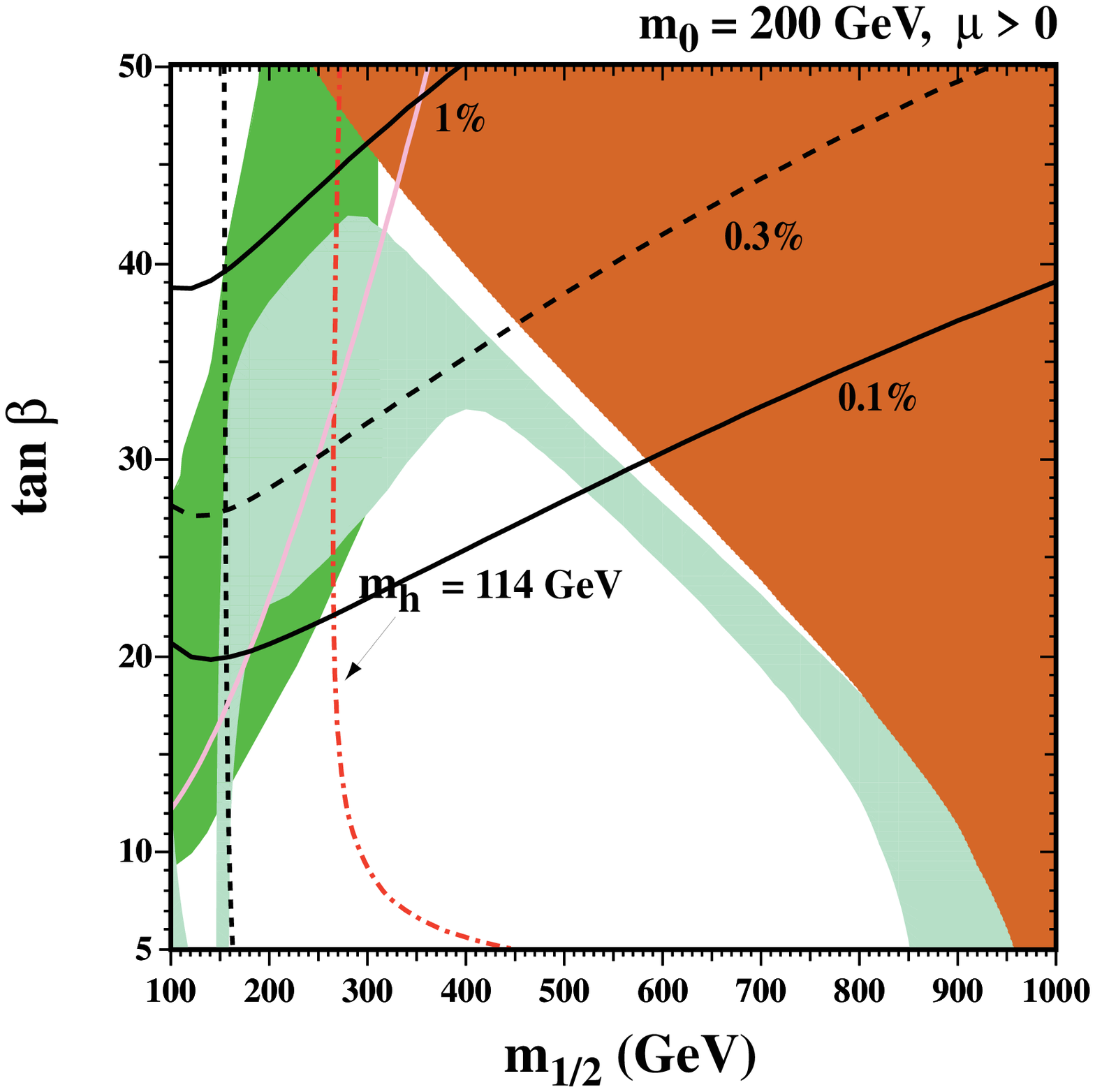 ,height=3.25in}       
\end{minipage}
\vskip .2in   
\caption{\label{fig3}
{\it The contours of $\mbox{A}_{\mbox{\small FB}}(\pi\tau^+\tau^-)$
(left panel) and $\mbox{A}_{\mbox{\small FB}}(K\mu^+\mu^-)$ (right
panel) for $A_0=0$, $m_0=200\ {\rm GeV}$ and $\mu>0$. All other curves and
shaded regions are taken from Fig. \protect\ref{fig1}c).}}
\end{figure}

In Fig. \ref{fig3}, we show the contours of $\mbox{A}_{\mbox{\small
FB}}(\pi\tau^+\tau^-)$ (left panel) and $\mbox{A}_{\mbox{\small
FB}}(K\mu^+\mu^-)$ (right panel) for $A_0=0$, $m_0=200\ {\rm GeV}$ and $\mu>0$. A
comparison  of the left panel with  panel c) of Fig. \ref{fig1} shows that there
is very little difference between $\pi \ell^+\ell^-$ and $K \ell^+\ell^-$
final states as far as the asymmetry is concerned. Indeed, as 
mentioned before, the difference between the asymmetries
is a measure of the  SU(3) flavor breaking or the different 
parameterizations of the associated form factors. Therefore, 
the similarity or dissimilarity of these two figures 
depends on how the hadronic effects are treated for the
kaon and pion final states. On the other hand, the comparison
between panel c) of Fig. \ref{fig1} and the right panel of Fig. \ref{fig3}
shows that the asymmetry is suppressed for the $\mu^+\mu^-$ 
final states. The asymmetry does not reach the $1\%$ level
in any corner of the allowed regions.

\begin{figure}
\vspace*{-0.25in}
%\hspace*{-0.7in}
\begin{minipage}{8in}

\epsfig{file=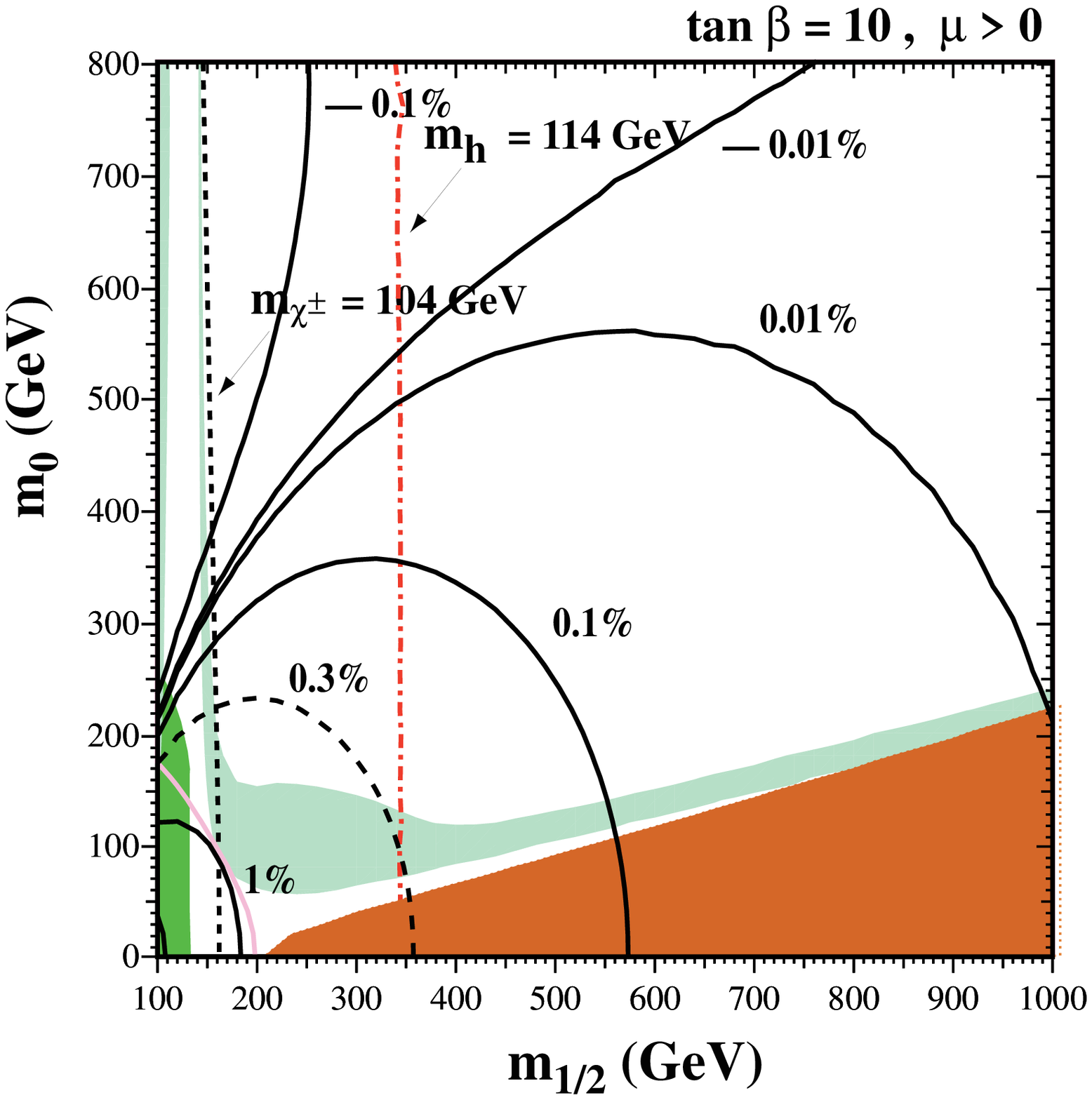,height=3.25in}
\epsfig{file=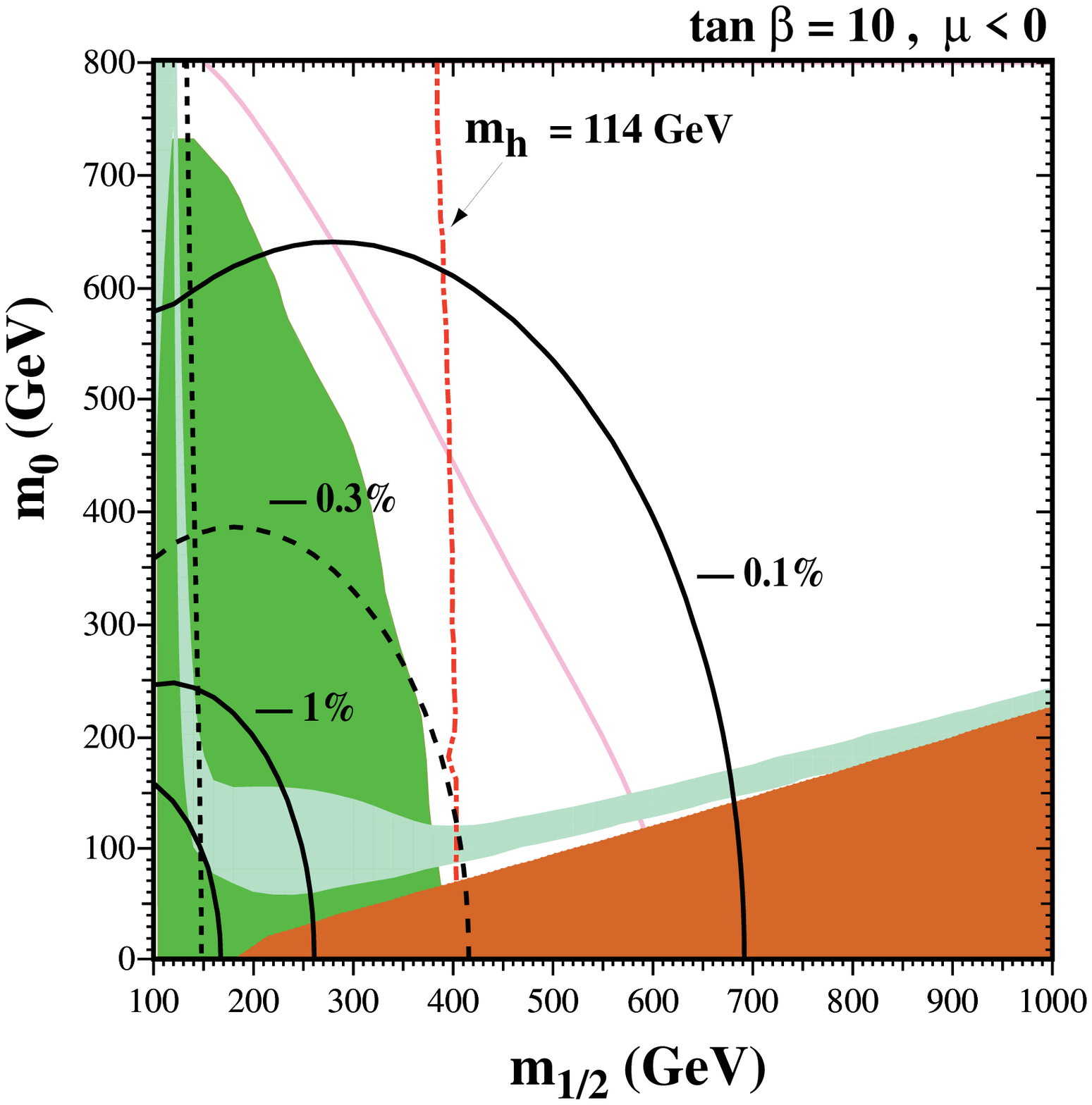,height=3.25in}
\end{minipage}
\begin{minipage}{8in}
\epsfig{file=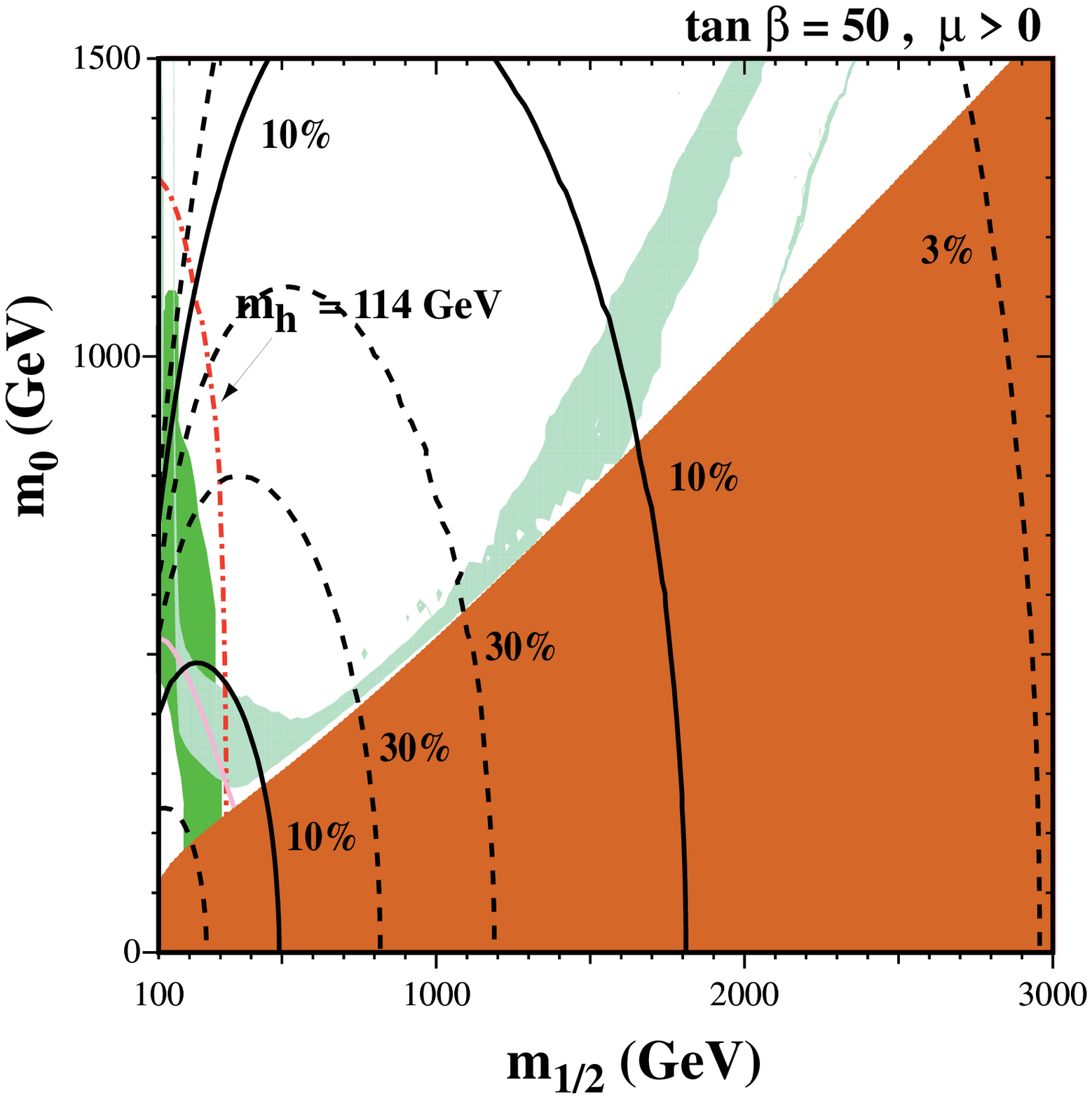,height=3.25in}
\epsfig{file=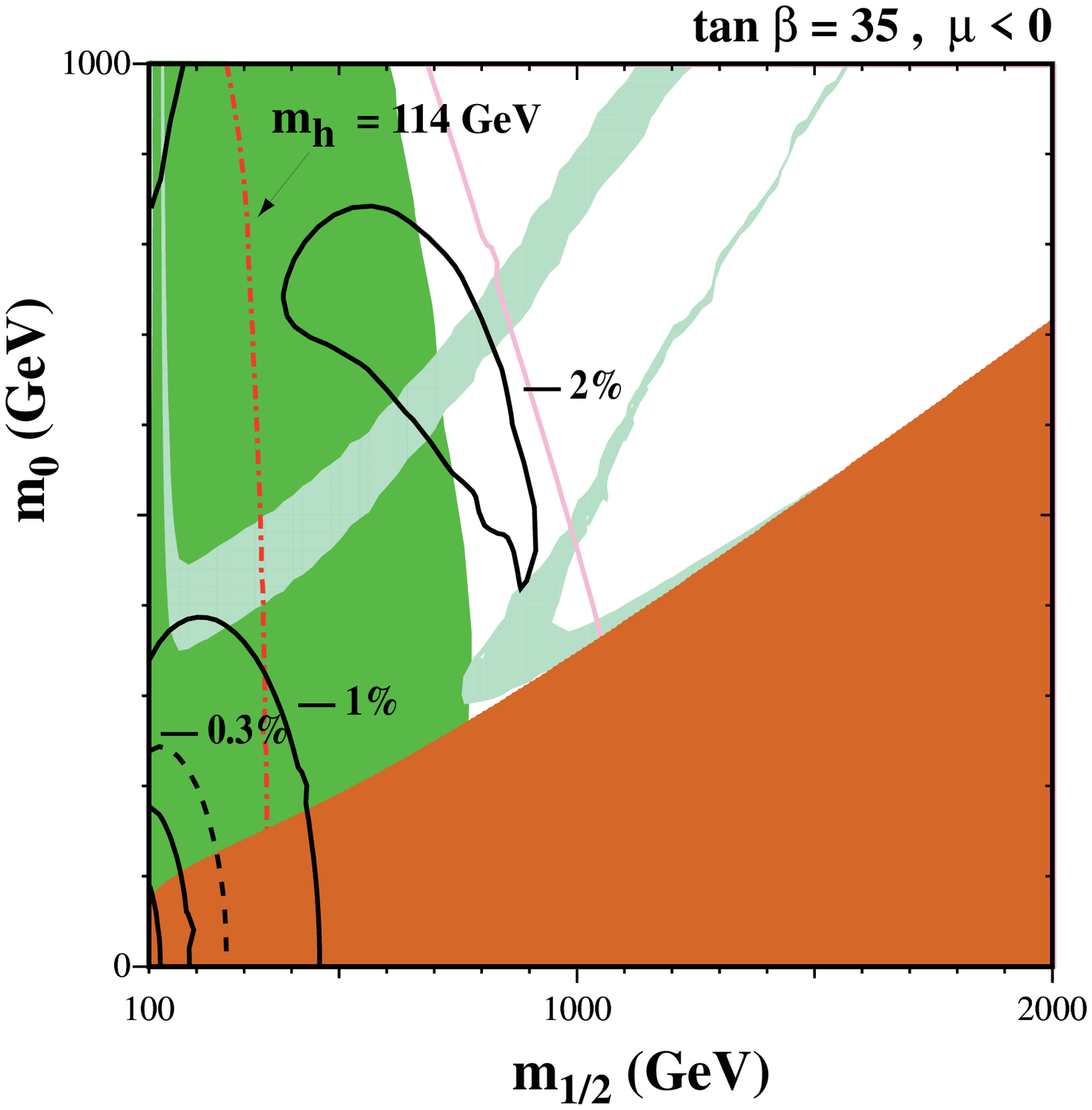,height=3.25in}
\end{minipage}
\vskip .2in
\caption{\label{fig4}
{\it The constant $\mbox{A}_{\mbox{\small FB}}(K\tau^+\tau^-)$ contours
in $m_{1/2}$--$m_0$ for $A_0=0$, and $\tan\beta=10$, $\mu>0$ (a);
$\tan\beta=10$, $\mu<0$ (b); $\tan\beta=50$, $\mu>0$ (c); and
$\tan\beta=35$, $\mu<0$ (d). Contours and shaded regions are as in Fig.
\protect\ref{fig1}}} 
\end{figure}

In Fig. \ref{fig4}a), we show the $\mbox{A}_{\mbox{\small FB}}(K\tau^+\tau^-)$
contours as projected onto  the $m_{1/2}$--$m_0$ plane for $A_0=0$,
$\tan\beta=10$,
$\mu>0$. The curves and shading in this figure are as in Fig. \ref{fig1}.
Seen clearly are the `bulk' and coannihilation cosmological regions. The latter
traces the border of the $\tilde \tau$ LSP region which is now found at
the lower right of the figure. The constraints from $b \to s \gamma$ and
$g-2$ only exclude a small portion of the parameter plane at the lower
left. In this figure, it is the area above the cosmological shaded region
which is excluded due to an excessive relic density.

 This figure illustrates the
dependence of the asymmetries on the chargino and  stop masses, and indeed, the
asymmetry changes sign at fairly large $m_0$ and small $m_{1/2}$. This effect
results in a sign change in ${\cal{C}}(m_b)$ due to the competition
between the stop and chargino masses. As Fig. \ref{fig4}a) makes clear,
the asymmetry is positive  and remains below the $0.3 \%$ level yielding
essentially no observable signal at all.

An example with $\tan \beta = 10 $ and $\mu < 0$ is shown in Fig. \ref{fig4}b).
While the cosmological region is similar to that in panel a),  the constraint
from $b \to s \gamma$ is significantly stronger, as is the constraint from $g-2$.
The latter imply that 
the asymmetry is never much larger than $-0.1\%$. As one can see, it is difficult
to get any observable effect  at low
values of  $\tan\beta$,for any sign of the $\mu$ parameter.

%\begin{figure}  
%\vspace*{-0.75in}
%\hspace*{-0.7in}
%\begin{minipage}{7.5in}
%\epsfig{file=asym3d.eps,height=3.25in}
%\epsfig{file=asym3c.eps,height=3.25in}
%\end{minipage}
%\vskip .2in
%\caption{\label{fig5}
%{\it The constant $\mbox{A}_{\mbox{\small FB}}(K\tau^+\tau^-)$ contours
%in $m_{1/2}$--$m_0$ plane for $A_0=0$, $\tan\beta=50$, $\mu>0$ 
%(left panel), and  $\tan\beta=35$, $\mu<0$ (right panel). In particular,
%the electroweak breaking condition prohibits larger values of $\tan\beta$
%for $\mu<0$.}}
%\end{figure}

In the lower two panels of Fig. \ref{fig4}, we show the contours of 
$\mbox{A}_{\mbox{\small FB}}(K\tau^+\tau^-)$  for larger values of $\tan \beta$.
As one can see, the `bulk' cosmological regions are pushed to higher values of 
$m_0$ and we also see the appearance of the `funnel' regions where the 
LSP relic density is primarily controlled by $H, A$ s-channel  annihilations.
In panel c), there are large positive 
asymmetries ($\gsim 10\%$) in a broad region extending from 
$(m_{1/2}, m_0)=(300, 300)\ {\rm GeV}$ all the way up 
to $(m_{1/2}, m_0)=(1.8, 1.1)\ {\rm TeV}$. The $B$--factory 
constraints only exclude a small region in the bottom left
corner bounded by  $m_0\lsim 200\ {\rm GeV}$ and 
$m_{1/2}\lsim 120\ {\rm GeV}$. Panel d) of Fig. \ref{fig4},
on the other hand, shows that, for negative $\mu$ and large $\tan \beta$,  the
allowed range of asymmetry is around $-1\%$ never reaching the $-2\%$ 
level mainly due to the $g_{\mu}-2$ constraint. There are 
three cosmologically preferred strips, the wider one located 
in the region  $(m_{1/2}, m_0)\gsim (900, 700)\ {\rm GeV}$ where
the asymmetry is at most $-1\%$. For $\mu<0$ the $B$--factory
constraints exclude two small regions bounded by $m_0\lsim 200\ 
{\rm GeV}$ and $m_{1/2}\lsim 280\ {\rm GeV}$, and
$m_0\lsim 200\  {\rm GeV}$ and $m_{1/2}\gsim 1940\ {\rm GeV}$
which both are already excluded by $\mbox{BR}(B\rightarrow X_s
\gamma)$ and cosmology.

\section{Summary}
We have discussed the forward--backward asymmetry of 
$B\rightarrow (K,\pi) \ell^+\ell^-$ decays which is
generated by the flavor--changing neutral current decays
mediated by the Higgs bosons. In addition to the
known properties $e.g.$ the smallness of the muon production
asymmetry compared to the $\tau$ production, the approximate 
independence of the asymmetry to the flavor of the 
final state meson, the reduction of the hadronic uncertainties
in the high dilepton mass region, $etc.$ we find that 
\begin{itemize}
\item The remarkable enhancement of the asymmetry 
is a unique implication of SUSY not found in the SM 
and its two--doublet version.

\item The regions of large asymmetry ($\gsim 10\%$) 
always require $\mu>0$, and when $\mu$ changes sign 
so does the asymmetry with an order of magnitude
suppression in its size.

\item The asymmetry in the decay channels with the lepton pair $\tau^+ \tau^-$
is significantly larger (approximately by the factor $m_\tau^2/m_\mu^2$) than in
the channels with $\mu^+\mu^-$. Thus in spite of a greater difficulty of
restoring the kinematics in the events with $\tau$ leptons, these may still be
advantageous for measuring the discussed asymmetry.

\item The asymmetry is not a monotonically increasing
function of $\tan\beta$ instead it is maximized 
at intermediate values above which the scalar FCNC
effects dominate and enhance the branching ratio,
and below which such FCNC are too weak to induce
an observable asymmetry.

\item Though $\mbox{BR}(B\rightarrow X_s \gamma)$ strongly
disfavors the negative values of $\mu$, the present
bounds from the muon $g-2$ measurement are much stronger, and it
typically renders the asymmetry unobservably small.

\item The cosmological constraints are generally 
very restrictive. When $m_0\sim 200$ GeV, 
$m_{1/2}\sim 400$ GeV, $\tan\beta\sim 30$ and $\mu>0$ 
there is a relatively wide region where the asymmetry 
is $\sim 30\%$ for $\tau$, and typically $\sim 0.5\%$ 
for muon final state.

\item The $B$--factory constraints are generally too
weak to distort the regions of observable asymmetry, 
and the regions excluded by them are already disfavoured by 
one or more of $\mbox{BR}(B\rightarrow X_s \gamma)$, 
$\widetilde{\tau}_{1}$ LSP and $g_{\mu}-2$. With increasing
statistics it is expected that the branching ratios
of semileptonic modes will be measured with better
accuracy so that, for instance, the allowed range 
in (\ref{belle}) will be narrowed. In this
case, there may be useful constraints on regions
of enhanced asymmetry.

\end{itemize}

\section{Acknowledgments}
This work is supported in part by the DOE grant DE-FG02-94ER40823.

\end{document}